\documentclass[prd,aps,twocolumn,a4paper,showkeys,nofootinbib]{revtex4-2}

\usepackage{graphicx,psfrag}
\usepackage{mathrsfs}
\usepackage{amsmath,amsfonts,amssymb}
\usepackage{multirow}
\usepackage{comment}
\usepackage{subfigure}
\usepackage{appendix}
\usepackage{hyperref}
\usepackage{ragged2e}

\usepackage{enumitem}

\newcommand{\be}{\begin{equation}}
\newcommand{\ee}{\end{equation}}
\newcommand{\bea}{\begin{eqnarray}}
\newcommand{\eea}{\end{eqnarray}}
\newcommand{\bel}{\begin{align}}
\newcommand{\eel}{\end{align}}

\def\e{{\rm e}}

\def\GMc2{{\rm G M_{\odot} c^{-2}}}

\def\kt2{\kappa^\text{T}_2}

\usepackage{pifont} 

\newcommand{\EFL}[1]{EFL{#1}} 
\usepackage{color}
\definecolor{cyan}{rgb}{0,0.9,0.9}
\definecolor{orange}{rgb}{0.9,0.5,0}
\definecolor{magenta}{rgb}{1,0,1}
\definecolor{purple}{rgb}{0.8,0.4,0.8}
\definecolor{gray}{rgb}{0.8242,0.8242,0.8242}

\begin{document}

\title{Entropy based flux limiting scheme for conservation laws}

\author{Georgios \surname{Doulis}$^{1,2}$}
\author{Sebastiano \surname{Bernuzzi}$^{2}$}
\author{Wolfgang \surname{Tichy}$^{3}$}

\affiliation{${}^1$Institut f{\"u}r Theoretische Physik, Goethe-Universit{\"a}t Frankfurt, 60438 Frankfurt am Main}
\affiliation{${}^2$Theoretisch-Physikalisches Institut, Friedrich-Schiller-Universit{\"a}t Jena, 07743 Jena}
\affiliation{${}^3$Department of Physics, Florida Atlantic University, Boca Raton, FL 33431}

\date{\today}

\begin{abstract}
The entropy based flux-limiting (EFL) scheme is a novel approach designed to accurately resolve shocks and discontinuities in special and general relativistic hydrodynamics. By adaptively adjusting the numerical fluxes, the EFL method mitigates oscillations and preserves smooth transition across discontinuities in shock-dominated flows.  Here, we extend the applicability of the EFL method beyond special/general relativistic hydrodynamics to scalar conservation laws and show how to treat systems without a thermodynamic entropy. This is an indication that the method has universal applicability to any system of partial differential equations that can be written in conservation form. We also present some further very challenging special/general relativistic hydrodynamics applications of the EFL method. 

\end{abstract}


\maketitle

\section{Introduction}

Binary neutron star (BNS) mergers are well-established sources of gravitational waves (GWs) \cite{TheLIGOScientific:2017qsa,GBM:2017lvd}. These events provide invaluable insights into astrophysical processes, compact object physics and the behaviour of matter under extreme conditions. Numerical relativity (NR) simulations are pivotal in reproducing merger waveforms with sufficient accuracy to enable reliable parameter estimation for BNS systems, shedding light on key properties such as mass ratios, tidal deformability and the equation of state (EoS) of neutron stars (NS).

Accurate numerical waveforms are indispensable for understanding the complex dynamics of BNS systems during the inspiral, merger and post-merger phases. Significant efforts have been devoted to developing high-resolution shock-capturing (HRSC) schemes \cite{Toro:1999} aiming at improving the accuracy and convergence properties of the numerical waveforms. Despite substantial progress, achieving convergence beyond second order remains a significant challenge for current HRSC schemes \cite{Bernuzzi:2012ci,Radice:2013hxh,Radice:2013xpa,Bernuzzi:2016pie}.

Among the various HRSC methods, we discuss here a specific family of techniques, known as flux-limiters \cite{Sweby:1984a}, that has demonstrated exceptional efficacy. Flux-limiters are designed to handle shock-like features robustly while suppressing spurious oscillations that can compromise numerical accuracy. These methods adaptively blend high- and low-order numerical fluxes, allowing for accurate resolution of shocks, rarefactions and other non-smooth features inherent in BNS simulations.

In \cite{Doulis:2022vkx}, we proposed a high-resolution HRSC scheme leveraging local entropy production in the design of an entropy based flux-limiter. The EFL scheme traces its origin to a method proposed in \cite{Guercilena:2016fdl}, where the use of entropy in the construction of a flux-limiter was first introduced. In \cite{Doulis:2022vkx} the EFL scheme was successfully used in several different special and general relativistic scenarios, including single and binary NS systems. Notably, the method achieved up to fourth-order convergence in the gravitational waveform phase, delivering high-fidelity results for both the dominant $(2,2)$-mode and the subdominant $(3,2)$- and $(4,4)$-modes.

In the present work we extend the applicability of the EFL method to scalar conservation laws without a thermodynamically defined entropy. To do so, an auxiliary entropy functional has to be defined \cite{Guermond:2011}, which can be used to flag the presence of non-smooth features in the solution space. This entropy can be used, like in \cite{Doulis:2022vkx}, to define the local weights entering the usual definition of the entropy flux-limiter. The weights combine an unfiltered high-order flux used in regions of smooth flow with a stable low- or high-order flux used in regions of non-smooth flow. 

The article is structured as follows. In \autoref{sec:EFL_method} we discuss the extension of the EFL method to scalar conservation laws. The first application of the EFL method to scalar conservation laws is presented in \autoref{sec:scalar_tests}. \autoref{sec:SRHD_tests} includes our results for some standard benchmark tests of special relativity and in \autoref{sec:SNS} the performance of our method is tested against three-dimensional general relativistic non-static/stationary single NS configurations. Finally, we conclude in \autoref{sec:conclusions}.

Throughout this work we use geometric units. We set $c = G = 1$ and the masses are expressed in terms of solar masses $M_\odot$.

\section{EFL Method}
\label{sec:EFL_method}

In the present work, the entropy flux limiting scheme developed in \cite{Doulis:2022vkx} is extended to systems without a thermodynamically defined entropy. Therefore, here, we consider any system of PDEs that can be written in conservation form: 
\begin{equation}
 \label{eq:conserv_PDE}
 \partial_t \textbf{Q} + \partial_i \textbf{F}^i(\textbf{Q}) = \textbf{S},
\end{equation}
where the summation is performed over the spatial dimensions, $\textbf{Q}$ is the vector of the conserved variables, $\textbf{F}^i$ the vector of the physical fluxes, and $\textbf{S}$ is the vector of the sources. From the plethora of system that can be written in the form \eqref{eq:conserv_PDE}, in the following, we are going to study the scalar transport and Burgers equations and the equations of special and general relativistic hydrodynamics \cite{Banyuls:1997zz}. 

The core concept of the EFL method lies in representing the numerical fluxes, derived from the spatial discretization of \eqref{eq:conserv_PDE}, as a weighted combination of two fluxes: an unstable high-order flux and a stable low- or high-order flux. The transition between these fluxes is governed by a locally computed weight, which is determined by the entropy production of the system. Essentially, the entropy is flagging regions of the computational domain where the flux calculation must switch from the unstable high-order scheme to the stable scheme to ensure numerical stability and accuracy.

For systems with thermodynamic entropy the implementation of the EFL method follows the exposition in \cite{Doulis:2022vkx}. In the following, we will show how to extend the EFL machinery of \cite{Doulis:2022vkx} to systems without a physically motivated definition of entropy, like the ones of \autoref{sec:scalar_tests}. 

Following the approach outlined in \cite{Doulis:2022vkx}, we begin by approximating the spatial derivative of the $x$-component, $\textbf{F}^x$, of the physical flux in \eqref{eq:conserv_PDE} using the conservative finite-difference scheme\footnote{We describe the method on one dimension. A multidimensional scheme can be constructed by calculating fluxes independently along each direction and summing them on the r.h.s.}
\begin{equation}
 \label{eq:spatial_disc}
  \partial_x F^x_i = \frac{\hat f_{i+1/2} - \hat f_{i-1/2}}{h},
\end{equation}
where $F^x$ is any one of the components of $\textbf{F}^x$ with $F^x_i=F^x(x_i)$, $\hat f_{i\pm1/2}$ are the numerical fluxes at the cell interfaces and $h$ is the spatial grid spacing.

Following the standard procedure \cite{Toro:1999}, we define the flux-limiter:
\begin{equation}
 \label{eq:num_flx_split}
  \hat f_{i\pm1/2} = \theta_{i\pm1/2} \hat f^{\,\mathrm{HO}}_{i\pm1/2} + 
  (1-\theta_{i\pm1/2}) \hat f^{\,\mathrm{LO}}_{i\pm1/2},
\end{equation}
where $f^{\,\mathrm{HO}}$ represents a high-order (HO) numerical flux, which is unstable but suitable for regions where the numerical solution remains smooth; $f^{\,\mathrm{LO}}$ denotes a high- or low-order (LO) stable numerical flux, applied in areas where the numerical solution exhibits non-smooth features; $\theta \in [0,1]$ is a continuous function blending these fluxes by determining the relative contribution of $f^{\,\mathrm{HO}}$ and $f^{\,\mathrm{LO}}$ during the dynamical evolution. To maintain consistency with previous studies \cite{Guercilena:2016fdl,Doulis:2022vkx}, we retain the notation LO, although we exclusively employ a stable high-order method to compute $f^{\,\mathrm{LO}}$ in the subsequent discussion. 

In particular, the flux $\hat f^{\,\mathrm{HO}}$ is built using the Rusanov Lax-Friedrichs flux-splitting technique and performing the reconstruction on the characteristic fields \cite{Mignone:2010br,Bernuzzi:2016pie}. A fifth-order central unfiltered stencil is always used for reconstruction. The flux $\hat f^{\,\mathrm{LO}}$ is approximated by the Local Lax-Friedrichs central scheme with reconstruction performed on the primitive variables \cite{Thierfelder:2011yi}. Primitive reconstruction here is performed with the fifth-order weighted-essentially-non-oscillatory finite difference schemes WENO5 \cite{Jiang:1996} and WENOZ \cite{Borges:2008a}.\footnote{To produce the results of \autoref{fig:bur_conv} the third-order convex-essentially-non-oscillatory (CENO3) algorithm \cite{Liu:1998,Zanna:2002} is used to perform primitive reconstruction.} 
    
The weight function $\theta$ is determined using the \textit{entropy production function} $\nu$, which is derived from the local entropy production. Specifically, the relationship between $\theta$ and $\nu$ is expressed as follows
\begin{equation}
 \label{eq:def_theta}
  \theta_{i\pm1/2} = 1- \frac{1}{2} (\nu_i + \nu_{i\pm1}).
\end{equation}   
Below, we summarise how to compute $\nu$ for systems with and without thermodynamically defined entropy.\\ 

\noindent\textbf{Systems with thermodynamic entropy}: 
For completeness, we repeat briefly here the exposition in \cite{Doulis:2022vkx}. To establish the relationship between $\nu$ and the entropy generated by the systems discussed  in Secs. \ref{sec:SRHD_tests} and \ref{sec:SNS}, we define the specific entropy (entropy per unit mass) for any piecewise polytropic EoS as
\begin{equation}
 \label{eq:spec_entrp}
  s = \ln \left( \frac{p}{\rho^\Gamma} \right),
\end{equation}
where the pressure is computed from the corresponding EoS. For a more general EoS, the specific entropy $s$ can be directly obtained from the EoS.

We define the entropy residual through the second law of thermodynamics as in~\cite{Guercilena:2016fdl}  
\begin{equation}
  \mathcal{R} = \nabla_\mu (s\,\rho\,u^\mu) \geq 0,
\end{equation}
which offers a quantitative estimate of the rate at which entropy is produced by the system. The expression above can be reformulated, see \cite{Guercilena:2016fdl}, in terms of the time and spatial derivatives of the specific entropy as follows
\begin{equation}
 \label{eq:phys_entrp_resid_GRR}
 \mathcal{R} = \frac{\rho W}{\alpha}
 \left(\partial_t s + (\alpha\, \upsilon^i - \beta^i) \partial_i s \right).
\end{equation}
In line with the reasoning presented in \cite{Doulis:2022vkx}, we omit the multiplicative factor $\frac{\rho W}{\alpha}$ and substitute $\mathcal{R}$ with
\begin{equation}
  \label{eq:phys_entrp_resid}
  R = \partial_t s + (\alpha\, \upsilon^i - \beta^i) \partial_i s.
\end{equation}

\noindent\textbf{Systems without thermodynamic entropy}: 
The systems of \autoref{sec:scalar_tests} do not have an EoS which can be used to define the entropy. For such systems \eqref{eq:conserv_PDE} takes the scalar form
\begin{equation}
 \label{eq:conserv_scalar_PDE}
 \partial_t Q + \partial_i F^i(Q) = S.
\end{equation}
Following \cite{Guermond:2008,Guermond:2011}, we define an entropy for \eqref{eq:conserv_scalar_PDE} by introducing the so-called entropy pair: $E(Q)$ and $\mathcal{F}^i(Q)$. The convex function $E$ is called entropy functional and $\mathcal{F}^i(Q)$ is the associated entropy flux given by the expression
\begin{equation}
 \label{eq:entr_flx}
 \mathcal{F}^i(Q) = \int \partial_Q E(Q)\, \partial_Q F^i(Q)\, dQ.
\end{equation}
The entropy pair can be used to define an entropy residual:
\begin{equation}
  \label{eq:scalar_entrp_resid}
  R = \partial_t E(Q) + \partial_i \mathcal{F}^i(Q).
\end{equation}
For convex physical fluxes $F^i$, like the ones of \autoref{sec:scalar_tests}, the entropy pair generated by the choice $E(Q) = \frac{1}{2}Q^2$ has been proven \cite{Guermond:2008,Guermond:2011} enough for our purposes here. With the above choice of the entropy functional, the entropy residual \eqref{eq:scalar_entrp_resid} can be straightforwardly computed for the systems of \autoref{sec:scalar_tests}.\\

Finally, we define the so-called \textit{entropy production function}, for systems with and without thermodynamic entropy, in terms of the rescaled entropy residual $R$ defined by \eqref{eq:phys_entrp_resid} or \eqref{eq:scalar_entrp_resid},
\begin{equation}
 \label{eq:nu_E}
  \nu_E = c_E |R|,
\end{equation}
where $c_E$ is a tunable constant used to scale the absolute value of $R$. In all our three-dimensional general relativistic simulations of Secs.~\ref{sec:SNS} we do not have to tune $c_E$, its value is set always to unity, i.e.\ $c_E=1$. For the one-dimensional simulations of Secs. \ref{sec:scalar_tests} and \ref{sec:SRHD_tests} the tunable constant $c_E$ is set either to 10 or 50 depending on the case studied.  

Considering that the parameter $\theta$ cannot exceed unity, we must impose a maximum value of  $\nu_{\rm max}=1$ on the entropy production function to ensure that the r.h.s of \eqref{eq:def_theta} remains within the range $[0,1]$. Therefore, the entropy production function in \eqref{eq:def_theta} is given by
\begin{equation}
 \label{eq:nu}
  \nu = \min\left[\nu_E, 1\right].
\end{equation}

The temporal evolution of the various systems discussed in this work is carried out using the general relativistic hydrodynamics finite differencing code BAM \cite{Brugmann:2008zz,Thierfelder:2011yi,Dietrich:2015iva,Bernuzzi:2016pie}. The EFL method has been integrated into BAM and is now part of its core infrastructure, see \cite{Doulis:2022vkx} for details.

\section{Scalar 1D tests}
\label{sec:scalar_tests}

\subsection{Transport equation with non-smooth initial data}
\label{sec:transport}

We start from the simplest one-dimensional scalar equation 
that describes the evolution of shocks:
\begin{equation}
 \label{eq:adv_eq}
  \partial_t u(t,x) + \partial_x u(t,x) = 0 \quad \mathrm{with} 
  \quad u(0,x)=u_0(x).
\end{equation}

In order to test the convergence properties of the EFL method 
we start with smooth initial data of the form $u_0(x) = \e^{-40 x^2}$. 
The exact solution is infinitely smooth and is deduced from 
the initial data: $u(t,x)=u_0(x-t)$. Convergence studies are 
performed on an equidistant grid with range $[-1,1]$ that is 
progressively refined with $n=2^k\times 200$ grid points, where 
$k=0,\ldots,5$. Different high-order RKp (with $3\leq p \leq 5$) 
schemes are used for the temporal evolution. Our convergence 
analysis indicates that optimal convergences $p$ is achieved 
in all cases independently of the scheme used for the LO flux. 
\begin{figure}[h]
 \centering
  \includegraphics[width=0.45\textwidth]{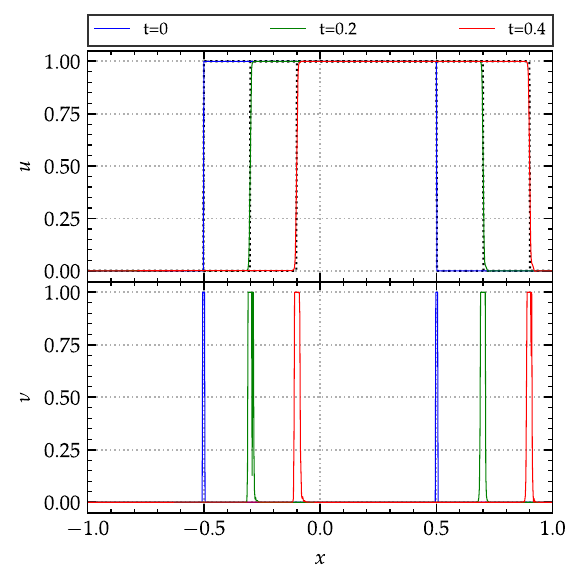}    
 \caption{Transport equation with non-smooth square wave 
 initial data. The numerical solution (top panel) and the 
 corresponding entropy production profile (bottom panel) 
 are depicted at different stages of the evolution. The 
 solution is computed with the WENOZ reconstruction scheme 
 on a grid of 800 points with resolution $\Delta x = 2.5
 \times 10^{-3}$. Dotted black lines depict the exact solution.}
  \label{fig:adv_eq_non_smooth}
\end{figure}

\begin{figure}[h]
  \centering
   \includegraphics[width=0.45\textwidth]{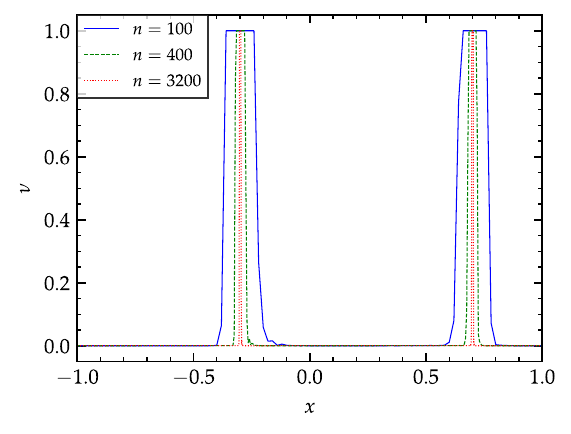}    
  \caption{Entropy production profiles for the transport 
  equation with increasing resolution. The profile of $\nu$ at 
  $t=0.2$ for a low $n=100$ (solid blue line), an intermediate 
  $n=400$ (dashed green line) and a high $n=3200$ (dotted red 
  line) resolution. The solution is computed with the WENOZ 
  reconstruction scheme.}
  \label{fig:adv_eq_nu}
\end{figure}

Next, we consider non-smooth square wave initial data: 
\begin{equation}
 \label{eq:adv_non_smooth_id}
  u_0(x)= 
  \begin{cases}
    1, & \text{if\,\, $-0.5 \leq x \leq 0.5$},\\
    0, & \text{otherwise}.
  \end{cases}
\end{equation} 
The grid set-up is the same as above. The initial data 
\eqref{eq:adv_non_smooth_id} is evolved with a RK4 routine 
and the Courant-Friedrichs-Lewy (CFL) condition is set to $0.125$. The WENOZ scheme 
is used for the LO flux and the value of the tunable constant 
is set to $c_E=10$. \autoref{fig:adv_eq_non_smooth} depicts 
the numerical solution $u$ of \eqref{eq:adv_eq} and the 
corresponding entropy production profile $\nu$ at different 
stages of the evolution on a grid composed by 800 grid-points 
($\Delta x=2.5\times 10^{-3}$). It is apparent that the 
EFL method locates successfully the position of the shocks and tracks them efficiently during the evolution. The fact 
that around the shocks of the square wave $\nu=1$ guarantees 
the use of the non-oscilatory LO flux there. Thus, the 
EFL method is able to suppress the emergence of oscillations 
around the vicinity of the shocks while achieving HO converenge 
on the rest of the computational domain where the solution 
is smooth. In addition, the behaviour of the entropy production 
profiles presented in \autoref{fig:adv_eq_nu} entails that 
with increasing resolution the entropy production gets better 
localised around the shocks and thus reduces the impact of 
the LO flux only to areas very close to the shocks.

Finally, \autoref{tab:adv_non_smooth_conv} contains the 
results of the convergence analysis for the numerical solution 
of \autoref{fig:adv_eq_non_smooth} at $t=0.2$. The observed 
convergence rates are in accordance with their theoretically 
expected values \cite{Toro:1999} of $1$ and $0.5$ for the norms $L_1$ and 
$L_2$, respectively. During the evolution the convergence 
rates are kept constant. Using lower that fifth order 
reconstruction schemes for the LO flux results in slightly 
smaller convergence rates, something expected as the lower 
accuracy of the LO flux drives down the overall accuracy 
of the entropy flux limiter. The use of other RK time 
integration routines, like RK3 and RK5, did not significantly 
affect the results of \autoref{tab:adv_non_smooth_conv}. 
\begin{table}[h]
 \centering    
 \caption{Convergence analysis of the numerical solution 
 of \autoref{fig:adv_eq_non_smooth} at $t = 0.2$. $L_1$ 
 and $L_2$ are normalised norms and the convergence rate 
 is calculated as the $\log_2$ of the ratio of two successive 
 normalised norms.}
   \begin{tabular}{ccccc}        
    \hline
    \hline
    n   & $L_1$   & Conv.& $L_2$   & Conv.\\
    \hline
    200 &\, $8.8\times 10^{-3}$\, & --    &\, $4.8\times 10^{-2}$\, &  --  \\
    400 &\, $5.1\times 10^{-3}$\, & 0.802 &\, $3.7\times 10^{-2}$\, & 0.398  \\
    800 &\, $2.9\times 10^{-3}$\, & 0.802 &\, $2.8\times 10^{-2}$\, & 0.396 \\
    1600&\, $1.7\times 10^{-3}$\, & 0.821 &\, $2.1\times 10^{-2}$\, & 0.397 \\
    3200&\, $9.3\times 10^{-4}$\, & 0.834 &\, $1.6\times 10^{-2}$\, & 0.401 \\
    6400&\, $5.2\times 10^{-4}$\, & 0.836 &\, $1.2\times 10^{-2}$\, & 0.412 \\
    \hline
    \hline
   \end{tabular}
 \label{tab:adv_non_smooth_conv}
\end{table}

\subsection{Burgers equation}

Next, we consider the inviscid Burgers equation in one-dimension:
\begin{equation*}
 \label{eq:bur_eq}
  \partial_t u(t,x) + u(t,x)\partial_x u(t,x) = 0 \quad 
  \mathrm{with} \quad u(0,x)=u_0(x).
\end{equation*}
This is the simplest non-linear scalar equation that develops 
shocks from originally smooth initial data:
\begin{equation}
 \label{eq:bur_smooth_id}
  u_0(x)= \e^{-40 x^2}. 
\end{equation} 
During the evolution the smooth initial profile becomes steeper 
and steeper and at around $t \simeq 0.1843$ a shock is formed.
\begin{figure}[h]
 \centering
  \includegraphics[width=0.45\textwidth]{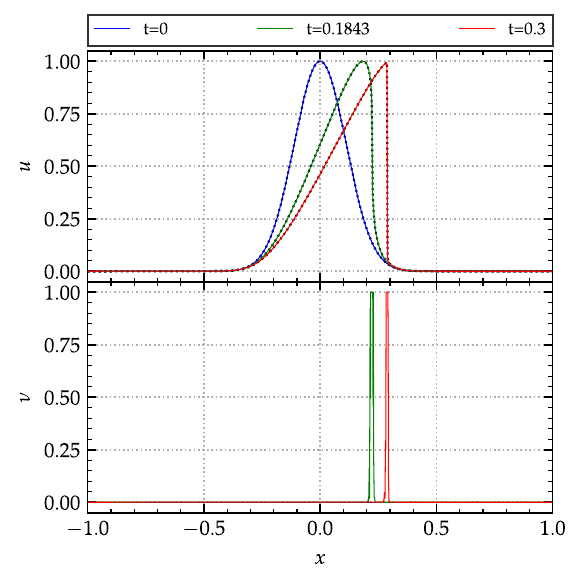}
  \caption{Burgers equation. The numerical solution (top panel) 
  and the corresponding entropy production profile (bottom panel) 
  are depicted at different stages of the evolution. The solution 
  is computed with the WENOZ reconstruction scheme on a grid of 
  1600 points with resolution $\Delta x = 1.25\times 10^{-3}$. 
  Dotted black lines depict the exact solution.}
  \label{fig:Burgers_eq}
\end{figure}

The exact solution is derived from the initial data 
\eqref{eq:bur_smooth_id} in the implicit form $u(t,x) = 
u_0(x-u t)$. The grid set-up is the same as in the case of 
the transport equation, see \autoref{sec:transport}. An RK4 
routine is used for the temporal evolution and the CFL condition 
is set to 0.125. The WENOZ scheme is used for the LO flux 
and the value of the tunable constant is set to $c_E=10$. 
\autoref{fig:Burgers_eq} depicts the numerical solution of 
Burgers equation (top panel) for the initial data \eqref{eq:bur_smooth_id} 
and the corresponding entropy production profile (bottom 
panel) at different stages of the evolution on a grid of 
1600 grid-points ($\Delta x=1.25\times 10^{-3}$). It is 
evident that the EFL method recognizes when the solution 
is smooth---the entropy production is zero---and, in addition, 
locates successfully the position of the emerging shock 
and tracks it efficiently during the evolution. It can be 
also confirmed by inspection that the EFL solutions are 
oscillation-free. The localisation of the entropy production 
profiles with increasing resolution first observed in 
\autoref{fig:adv_eq_nu} is also observed here.
\begin{figure}[h]
  \centering
   \includegraphics[width=0.45\textwidth]{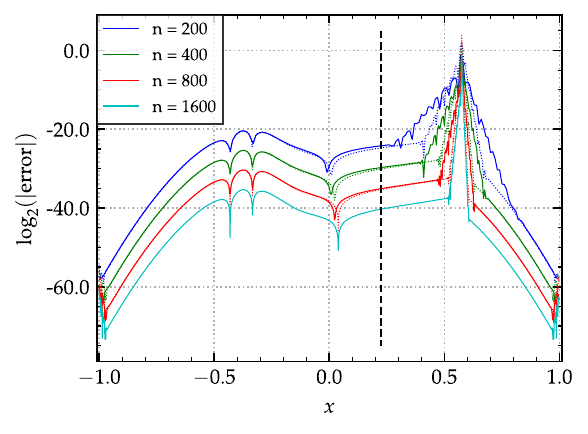}    
  \caption{Convergence to zero with increasing resolution 
  of the error between the numerical and the exact solution 
  of Burgers equation at $t=1$. The vertical dashed black 
  line depicts the position at which the shock is first 
  formed. Dotted lines show results scaled to fifth-order.}
  \label{fig:bur_dom_conv}
\end{figure} 

\autoref{fig:bur_dom_conv} depicts the numerical error with 
increasing resolution over the whole computational domain at 
$t=1$. Optimal fifth-order convergence is observed at the 
parts of the domain where the solution is smooth---dotted 
lines show results scaled to fifth-order. As expected close 
to the location of the shock the convergence drops to first-order. 
The position of the right-moving shock can be easily identified 
by the lower order convergence spike at $x \approx 0.6$. It 
is clear that the movement of the shock through the computational 
grid does not contaminate the rest of the numerical domain 
with its low order convergence. It is evident that the points 
of the computational domain affected by the shock recover the 
optimal fifth-order convergence almost instantly. Therefore, 
lower converegence is a local characteristic of our solution 
confined to the neighborhood of the shock.  

\begin{figure}[h]
  \centering
   \includegraphics[width=0.45\textwidth]{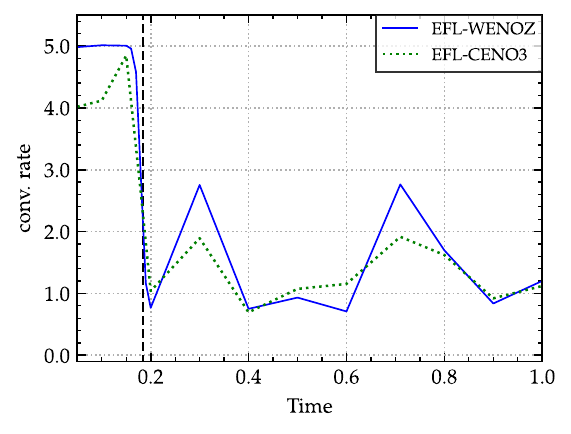}    
  \caption{Convergence rates with respect to the $L_1$-norm 
  for the numerical solution of Burgers equation. The 
  behaviour of the convergence rates with time for two 
  different LO schemes is shown: \EFL{-WENOZ} (solid 
  blue line) and \EFL{-CENO3} (dotted green line). The 
  vertical dashed black line depicts the instant at which 
  the shock is first formed.}
  \label{fig:bur_conv}
\end{figure}

In \autoref{fig:bur_conv} the convergence rates with 
respect to the $L_1$-norm for the numerical solution of 
Burgers equation are computed at different stages of the 
evolution for two different choices of the LO flux. At 
early stages of the evolution the convergence rates maintain 
their optimal high-order values. When the shock starts 
forming---see vertical dashed black line---the rates decrease 
rapidly and, after the shock has been fully formed, reach 
a plateau at around 1. Notice that in the smooth region 
before the formation of the shock, the use of a lower than 
fifth-order reconstruction scheme, i.e. CENO3, leads to 
slightly smaller convergence rates. The use of other RK 
routines does not alter the convergence rates of 
\autoref{fig:bur_conv} significantly.

\section{Special relativistic 1D tests}
\label{sec:SRHD_tests}

\subsection{Simple wave}

\begin{figure}[h]
 \centering 
  \includegraphics[width=0.45\textwidth]{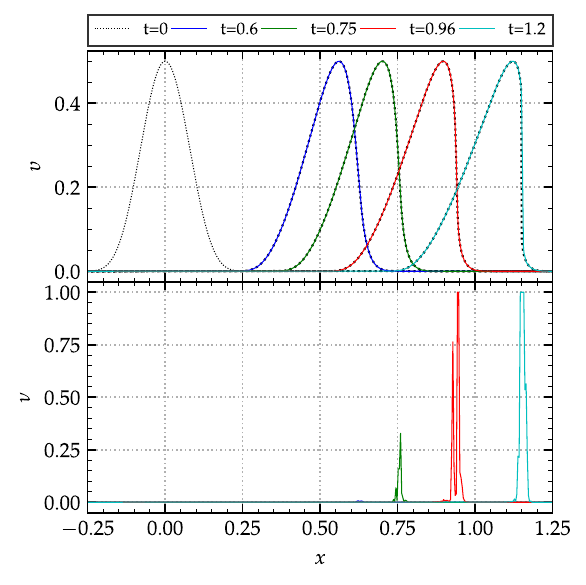}
   \caption{Simple wave. The numerical solution of the 
   velocity (top panel) and the corresponding entropy 
   production profile (bottom panel) are depicted at 
   different stages of the evolution. The solution is 
   computed with the WENOZ reconstruction scheme on a 
   grid of 1600 points with resolution $\Delta x = 1.875
   \times 10^{-3}$. Dotted black lines depict the exact 
   solution.}
 \label{fig:simple_wave_vx_nu}
\end{figure}

\begin{figure*}[t]
  \centering
  \begin{tabular}[c]{cc} 
   \includegraphics[width=0.43\textwidth]{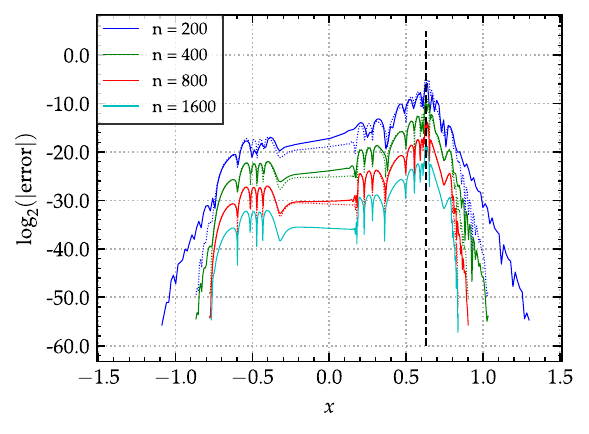}
   \put(-105,142){t=0.6}  
   \includegraphics[width=0.43\textwidth]{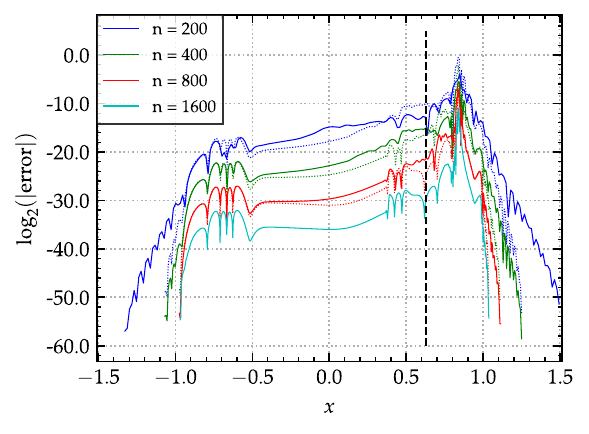}
   \put(-105,142){t=0.84} 
   \vspace{-2mm}
   \\
   \includegraphics[width=0.43\textwidth]{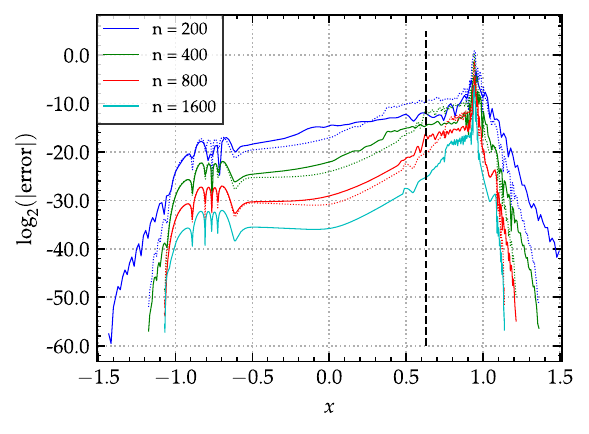}
   \put(-105,142){t=0.96} 
   \includegraphics[width=0.43\textwidth]{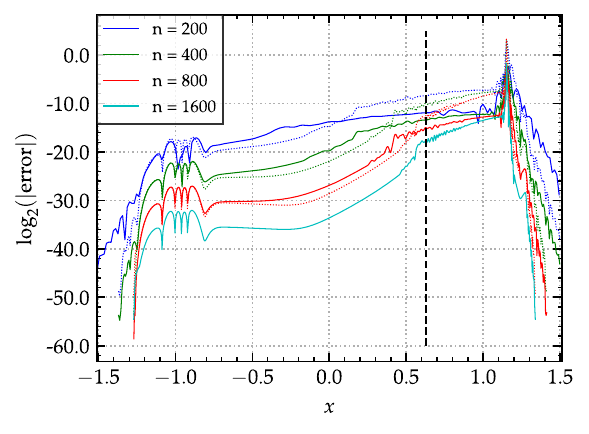}
   \put(-105,142){t=1.2} 
  \end{tabular}    
  \caption{Behaviour of the simple wave's numerical error 
  with increasing resolution at different instances of 
  the evolution. The vertical dashed black line depicts 
  the position at which the shock is first formed. Dotted 
  lines show results scaled to fifth-order.}
  \label{fig:simple_wave_dom_conv}
\end{figure*}

Relativistic simple waves are exact analytical solutions 
of special-relativistic hydrodynamics (SRHD) that 
demonstrate some of the main features of non-linear 
wave propagation, such as steepening, breaking, and shock 
formation. Simple wave solutions can be written analytically 
in an implicit form using the method of characteristics. 
The study of relativistic simple waves sheds light on the
problem of relativistic shock formation due to non-linear 
steepening. Like Burgers equation the relativistic simple 
wave starts off from smooth initial data and at some point 
of its evolution develops a shock. These analytic solutions 
have been discussed in \cite{Liang:1977a,Anile:1990a}. Here, 
we use the simple wave described in \cite{Bernuzzi:2016pie}, 
therein the initial velocity profile is of the form
\begin{equation}
 \label{eq:v_profile}
  \upsilon = a\, \Theta(|x| - X) \sin^6 \left( \frac{\pi}{2} \left( 
  \frac{x}{X} -1 \right) \right),
\end{equation} 
where $\Theta(x)$ is the Heaviside function, $a = 0.5$ 
and $X = 0.3$. For the construction of the initial data 
a polytropic EoS, $p=K\rho^\gamma$, with $K=100$ and 
$\gamma=5/3$ is employed; this data is then evolved using 
a $\Gamma$-law EoS with $\Gamma=5/3$. 
During the evolution the smooth initial profiles of all 
primitive variables become steeper and steeper and at 
around $t_{sh} \simeq 0.63$ they form a shock \cite{Liang:1977a}. 
Numerical solutions are computed on the one-dimensional 
domain $x \in [-1.5, 1.5]$ with the RK4 time-integrator 
and a CFL factor of 0.125. The tunable constant is set 
to $c_E=10$ and the LO flux is evaluated with the WENOZ 
reconstruction scheme. 

\begin{figure}[t]
 \centering 
  \includegraphics[width=0.45\textwidth]{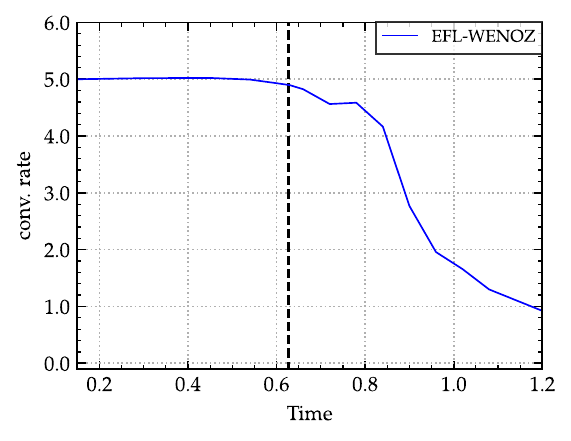}
   \caption{Convergence rates of the simple wave with 
   respect to the $L_1$-norm. The behaviour of the 
   convergence rates with time for the LO scheme 
   EFL-WENOZ is shown. The vertical dashed black line 
   depicts the instant at which the shock is first 
   formed.}
 \label{fig:simple_wave_conv}
\end{figure}

The top panel of \autoref{fig:simple_wave_vx_nu} illustrates 
the evolution of the simple wave from its initial smooth 
configuration \eqref{eq:v_profile} up to the instance 
$t = 1.2$ when the shock has been fully formed. The 
solution is computed on a grid composed of 1600 points 
($\Delta x = 1.875\times 10^{-3}$). By inspection, the 
numerical solutions reproduce the correct physics even 
after the formation of the shock $t \geq t_{sh}$. No 
spurious oscillations are visible on the solutions, a 
sign that shows the effectiveness of the EFL method in 
suppressing oscillatory effects. The bottom panel of 
\autoref{fig:simple_wave_vx_nu} depicts the corresponding 
profiles of the entropy production function $\nu$. 
Clearly, the EFL method successfully locates the initial 
formation of the shock at $t = t_{sh}$ and effectively 
tracks its location at later times $t \geq t_{sh}$. 
During the ``smooth period'' $t < t_{sh}$, the entropy 
production is, as expected, trivial. Notice that shortly 
after the formation of the shock, i.e. at $t = 0.75$, 
the solution is oscillation-free although the contribution 
of the stable LO flux in the hybrid numerical flux 
\eqref{eq:num_flx_split} is only $\sim 33\%$. This fact 
points out that the EFL method can guarantee the smoothness 
of the solutions for some time after the shock formation 
without switching entirely to the LO flux. This property 
of the EFL method has a direct effect on the behaviour 
of the convergence rates of \autoref{fig:simple_wave_conv}. 
As expected, optimal fifth-order convergence is maintained 
up to the formation of the shock---denoted here by a 
vertical dashed line. With the shock formed the convergence 
of our numerical solution decreases slowly up to roughly 
$t \sim 0.85$---notice that the convergence rates are 
still higher than fourth-order. Beyond this point the 
convergence decreases rapidly and reaches a plateau 
at around 1. Clearly, there is a direct connection 
between the increasing use of the LO flux and the 
decreasing order of convergence. 

\autoref{fig:simple_wave_dom_conv} depicts the absolute 
error between the numerical and the exact solution with 
increasing resolution at different stages of the evolution. 
For times $t < t_{sh}$ before the formation of the shock 
the numerical error decreases to zero with the expected 
fifth-order convergence, see top-left panel of 
\autoref{fig:simple_wave_dom_conv}. With the formation 
of the shock at $t = t_{sh}$ the expected lower-order 
convergence spike appears. Shortly after, the numerical 
error convergences to zero with fifth-order over the 
whole computational domain except at the location of 
the shock where the convergence drops to first-order, 
see top-right panel of \autoref{fig:simple_wave_dom_conv}. 
Notice that, at this stage, the convergence of the parts 
of the computational domain impacted by the shock is 
quickly restored to fifth-order. At late times, see 
bottom panels of \autoref{fig:simple_wave_dom_conv}, 
this is not anymore possible and the low-order convergence 
spreads through the computational grid at the opposite 
direction of the movement of the shock. This is in 
stark contrast with the behaviour of the numerical 
error of Burgers equation, see \autoref{fig:bur_dom_conv}. 
The coupled non-linear nature of the SRHD equations 
is responsible for this difference as the EFL method 
operates in both cases in a similar fashion: the LO 
flux dominates the hybrid numerical flux \eqref{eq:num_flx_split} 
and its use has been restricted around the location of 
the shock.

\subsection{Sod shock-tube} 

\begin{figure}[b]
 \centering 
  \includegraphics[width=0.45\textwidth]{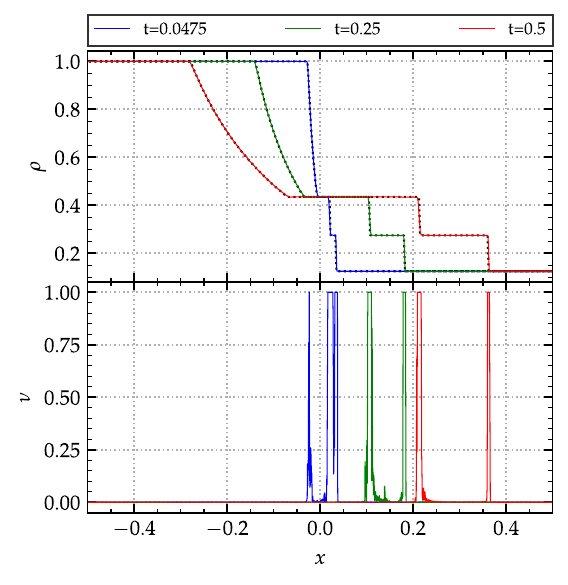}
   \caption{Sod shock-tube. Profiles of the rest-mass 
   density (top panel) and of the corresponding entropy 
   production (bottom panel) at different stages of 
   the evolution. The solution is computed with the 
   WENO5 reconstruction scheme on a grid of 1600 points 
   with resolution $\Delta x = 6.25 \times 10^{-4}$. 
   Dotted black lines depict the exact solution.}
 \label{fig:sod_vx_nu}
\end{figure}

We move on now to a standard benchmark Riemann problem 
used in SRHD: the relativistic version of Sod's shock-tube 
\cite{Sod:1978}. For a simple ideal fluid EoS with adiabatic index $\Gamma 
= 1.4$, the discontinuous initial data for the pressure 
$p$, the rest-mass density $\rho$, the velocity $\upsilon$, 
and the specific energy $\epsilon$ are the following
\begin{equation}
 \label{eq:Sod_id}
  \begin{aligned}
   ( p_L, \rho_L, \upsilon_L, \epsilon_L ) &= (1,   1,     0, 2.5 ), \\ 
   ( p_R, \rho_R, \upsilon_R, \epsilon_R ) &= (0.1, 0.125, 0, 2 ).
  \end{aligned}
\end{equation}
The initial discontinuity at $x=0$ splits during the 
evolution into a shock wave followed by a contact 
discontinuity, both travelling to the right, and a 
rarefaction wave travelling to the left. The numerical 
solutions are computed on the one-dimensional grid 
$x \in [-0.5, 0.5]$ with the RK4 scheme and a CFL 
factor of 0.25. The tunable constant is set to $c_E=50$ 
and the WENO5 reconstruction scheme is used to evaluate 
the LO flux. 

The top panel of \autoref{fig:sod_vx_nu} shows how 
the initial discontinuity at $x=0$ evolves with time 
on a grid of 1600 points ($\Delta x = 6.25\times 10^{-4}$). 
Even at early stages of the evolution the three 
characteristic features of Sod's shock-tube are clearly 
visible: the left-moving rarefaction wave and the 
right-moving shock and contact discontinuities. It 
is evident that all the features of the Sod shock-tube 
are reproduced quite accurately and oscillatory effects 
are absent from the solutions. At the bottom panel 
of \autoref{fig:sod_vx_nu} the corresponding entropy 
production profiles are shown. Again the EFL method 
locates and tracks the right-moving discontinuities 
extremely accurately. Notice that the entropy production 
peaks are very well localised around the position of 
the discontinuities, restricting in this way the use 
of the stable LO flux only to these problematic regions 
and enabling the use of the HO flux in the rest of 
the computational domain where the solution is smooth. 
At early stages of the evolutions a third entropy 
production peak related to the rarefaction wave is 
clearly visible. The initial very steep slope of the 
rarefaction triggers the production of this peak. With 
time, as the rarefaction develops, the amplitude of 
the entropy production peak reduces and tends to zero 
when the rarefaction wave has been fully developed. 
This comprises another manifestation of the sensitivity 
and accuracy of the EFL method.  

\begin{figure}[h]
 \centering 
  \includegraphics[width=0.45\textwidth]{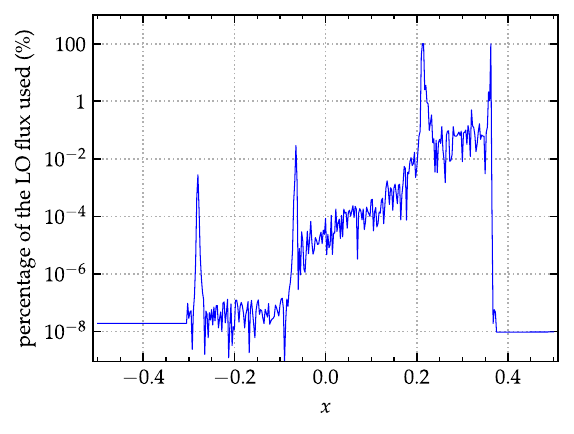}
   \caption{Percentage of the stable LO flux used at 
   $t=0.5$ over the whole computational domain.}
 \label{fig:sod_lo_perc}
\end{figure}

A more quantitative picture of the use of the LO flux 
at each point of the computational domain is given in 
\autoref{fig:sod_lo_perc}. As mentioned in \autoref{sec:EFL_method} 
the entropy production takes values in the interval 
$\nu \in [0,1]$. In the extreme cases of $\nu=0$ or 
$\nu=1$ the HO or LO flux is exclusively used by the 
EFL method. For any other value of $\nu$ the resulting 
numerical flux \eqref{eq:num_flx_split} is a mix of 
these two fluxes. \autoref{fig:sod_lo_perc} shows 
exactly that: the percentage of each flux used at 
each point of the computational domain at $t=0.5$. 
As expected the stable LO flux is exclusively used 
at the locations of the contact and shock discontinuities, 
in between them the use of the LO flux drops drastically 
to $\sim 0.5\%$ and at any other point of the computational 
domain does not exceed $0.01\%$. Notice that the 
positions of the head and tail of the rarefaction 
wave are clearly indicated by the two $0.01\%$ 
entropy production peaks---the EFL method is so 
sensitive that recognises the slight non-smoothness 
of these two features compared to their neighbouring 
points. 

Similar results we get also for the other two 
standard SRHD shock-tube tests: the relativistic 
blast wave 1 and 2 described in \cite{Marti:1999wi}.

\section{Single star evolutions}
\label{sec:SNS}

We proceed further by applying the EFL method to three-dimensional general relativistic single NS spacetimes. In addition to the static TOV and rotating NS tests presented in \cite{Doulis:2022vkx}, we study here a couple of more involved single NS spacetimes that unveil the ability of the EFL method to cope extremely well with non static/stationary spacetimes. The treatment of the rapidly declining gradient of the hydrodynamical variables at the surface of the NS remains also here the most challenging feature of our simulations. In addition, the non static/stationary nature of these spacetimes increases further the level of complexity we encounter. Our results are compared with those of \cite{Font:1998hf,Font:2001ew,CorderoCarrion:2008nf,Thierfelder:2011yi}. The EFL method is also compared to the best performing schemes currently implemented in BAM: i) a second-order scheme (LLF-WENOZ) that uses the LLF scheme for the fluxes and WENOZ for primitive reconstruction \cite{Thierfelder:2011yi} and ii) a ``hybrid'' algorithm (HO-LLF-WENOZ) that employs the high-order HO-WENOZ scheme above a certain density threshold $\rho_\mathrm{hyb}$ and switches to the standard second-order LLF-WENOZ method below $\rho_\mathrm{hyb}$ \cite{Bernuzzi:2016pie}. 

In the following, we evolve a stable TOV star that moves along the $x$-direction \cite{Font:1998hf,Thierfelder:2011yi} and an unstable migrating NS \cite{Font:2001ew,CorderoCarrion:2008nf,Thierfelder:2011yi}, both in a dynamically evolved spacetime. The NS matter is here described by a $\Gamma$-law EoS with $\Gamma=2$. The grid is composed of five fixed refinement levels. Simulations are performed at resolutions $n=(64,96,128)$ points leading to a grid spacing $h$ that depends on the specific setting of the NS under investigation. For each NS configuration the resolution is explicitly given in \autoref{tab:SNS_grid}. It is ensured that the NS is entirely covered by the finest box at any given resolution. Radiative (absorbing) boundary conditions are used for all single star simulations.

\begin{table}[h]
 \centering   
 \caption{Grid configurations of single star simulations. Columns (left to right): name of simulation, $L$: number of fixed refinement levels, $n$: number of points per direction, $h_{L-1}$: resolution per direction in the finest level $l=L-1$, $h_0$: resolution per direction in the coarsest level $l = 0$.}
   \begin{tabular}{c|cccccc}        
    \hline
    \hline
    Name & $L$ & $n$ & $h_{L-1}$ & $h_0$ \\
    \hline
    \multirow{3}{4.5em}{TOV\textsubscript{boost}} 
    & 5  &  64  & 0.500  & 8.000  \\
    & 5  &  96  & 0.333  & 5.333   \\
    & 5  &  128 & 0.250  & 4.000   \\
    \hline
    \multirow{3}{4.5em}{TOV\textsubscript{mig}} 
    & 5  &  64  & 0.125  & 2.000  \\
    & 5  &  96  & 0.083  & 1.333   \\
    & 5  &  128 & 0.063  & 1.000   \\
    \hline
    \hline
   \end{tabular}  
 \label{tab:SNS_grid}
\end{table}

\subsection{Boosted TOV star}
\label{sec:boost_ideal_dyn_tov}

The Tolmann-Oppenheimer-Volkoff (TOV) initial data used here is constructed using a $\Gamma=2$ polytrope model with $K=100$, gravitational mass $M=1.4$, baryonic mass $M_b=1.506$ and central rest-mass density $\rho_c=1.28 \times10^{-3}$. The resulting TOV star is boosted in the $x$-direction at a speed of $v = 0.5$ corresponding to a Lorentz factor of $W = 1.15$. The spacetime is dynamically evolved and the BSSNOK scheme is used for the evolution of the metric.

\begin{figure}[h]
 \centering 
  \includegraphics[width=0.45\textwidth]{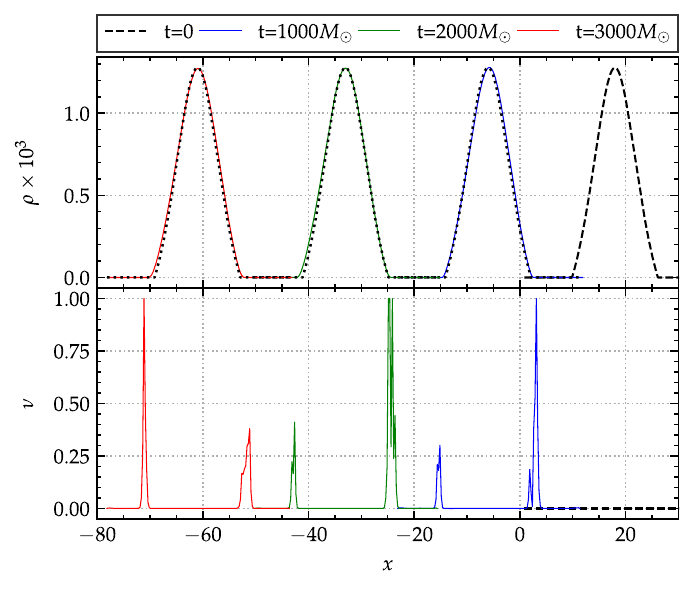}
   \caption{Boosted TOV star. Profiles of the rest-mass density (top panel) and of the corresponding entropy production (bottom panel) at different instances of the movement of the TOV star along the $x$-direction. The solution is computed with the WENOZ reconstruction scheme and $n=128$. Dotted black lines depict the exact solution.}
   \label{fig:1d_rho_nu_dyn_ideal_boost}
\end{figure}

The one-dimensional profiles of the rest-mass density (top panel) and entropy production function $\nu$ (bottom panel) are depicted in \autoref{fig:1d_rho_nu_dyn_ideal_boost}. The top panel shows the evolution of the TOV star along the $x$-direction from its initial position (dashed black lines) up to $t=3000M_\odot$. The numerical solution reproduces quite accurately the exact one (dotted black lines) even at very late times. At the bottom panel the corresponding entropy production profiles are depicted. As expected, a local peak of the entropy production function $\nu$ is observed around the surface of the TOV star. There, the gradient of the hydrodynamical variables experiences a violent decline which leads to the production of large values of $\nu$. In the interior of the NS the entropy production function $\nu$ is vanishing. It is evident from \autoref{fig:1d_rho_nu_dyn_ideal_boost} that the EFL scheme locates and tracks extremely accurately the surface of the star during its movement along the $x$-axis. This, in turn, triggers the use of the stable numerical flux in the neighbourhood of the surface. The use of the stable scheme around the surface of the NS guarantees its long term stability.

\begin{figure}[h]
 \centering 
  \includegraphics[width=0.45\textwidth]{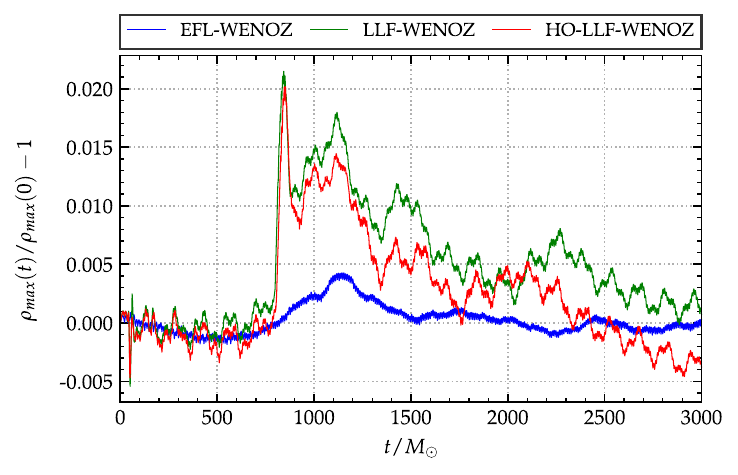}
   \caption{Central rest-mass density evolution of a boosted TOV star in a dynamical spacetime with $n=128$. Beside the EFL-WENOZ scheme, the LLF-WENOZ and HO-LLF-WENOZ schemes are also shown.}
   \label{fig:rho_max_dyn_ideal_boost}
\end{figure}

After securing the proper flagging of the problematic regions and the correct implementation of the EFL method, we proceed further by checking its performance during the evolution of the TOV star by monitoring the dynamical behaviour of the central rest-mass density. The oscillation of the central rest-mass density $\rho_{\rm max}$ for the EFL method is presented, together with the results for the LLF-WENOZ and HO-LLF-WENOZ methods \cite{Bernuzzi:2016pie}, in \autoref{fig:rho_max_dyn_ideal_boost}. Clearly, the EFL method is performing better than the LLF-WENOZ and HO-LLF-WENOZ schemes as the profile of the central rest-mass density is more stable over time and the amplitude of the oscillations is smaller. 

\begin{figure}[h]
 \centering 
  \includegraphics[width=0.45\textwidth]{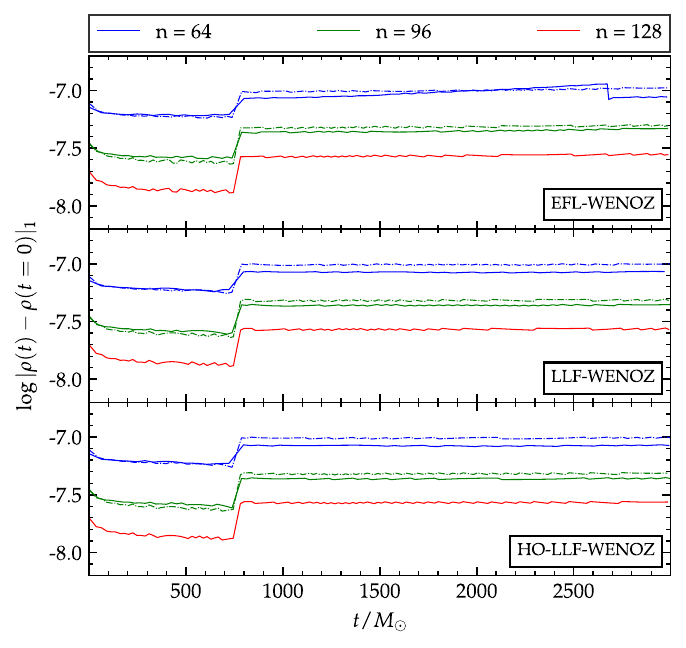}
   \caption{Evolution of the $L_1$ distance $||\rho(t)-\rho(0)||_1$ of a boosted TOV star in a dynamically evolved spacetime. The EFL-WENOZ scheme is compared to the LLF-WENOZ and HO-LLF-WENOZ schemes. Dashed lines show results scaled to second order.}
   \label{fig:ideal_l1_rho_dyn_boost}
\end{figure}

The non-dispersive character of the boosted TOV star enables us to check the convergence properties of the EFL method as the exact solution can be read off from the initial data. Here we consider the $L_1$-norm of the difference between the three-dimensional evolution profile of the rest-mass density and the corresponding exact solution (initial data) and study its behaviour with time. The $L_1$-distance from the exact solution for the three schemes used here are depicted in \autoref{fig:ideal_l1_rho_dyn_boost}. The convergence rate of all the schemes considered is approximately second-order in agreement with the result of \cite{Thierfelder:2011yi} and the fact that the error at the stellar surface dominates the evolution.

\subsection{Migrating neutron star}
\label{sec:migration_ideal_dyn_tov}

We turn now to the study of migrating neutron stars and focus on the migration of unstable neutron stars to stable ones. This is a quite challenging test as during the process of migration oscillations of large amplitude are produced by the pulsating NS. The rapid  periodic expansion and contraction of the surface of the NS is the ideal arena to test the EFL method's performance in an extreme dynamical setting where the location of the surface is constantly changing in a non-trivial way. 

The unstable initial configuration we consider here is constructed using a $\Gamma=2$ polytrope model with $K=100$, gravitational mass $M=1.447$, baryonic mass $M_b=1.535$ and central rest-mass density $\rho_c=7.993\times10^{-3}$ that is larger than the central rest-mass density of the stable model with the maximum possible mass. Therefore the  NS is initially in an unstable equilibrium. Numerical truncation errors cause the NS to move away from this initial equilibrium and to expand, evolving gradually to smaller central rest-mass densities, until it reaches a new stable NS configuration. 

The star is evolved with the $\Gamma$-law EoS and the metric components with the BSSNOK scheme. The spacetime is dynamically evolved. 

\begin{figure}[h]
 \centering 
  \includegraphics[width=0.45\textwidth]{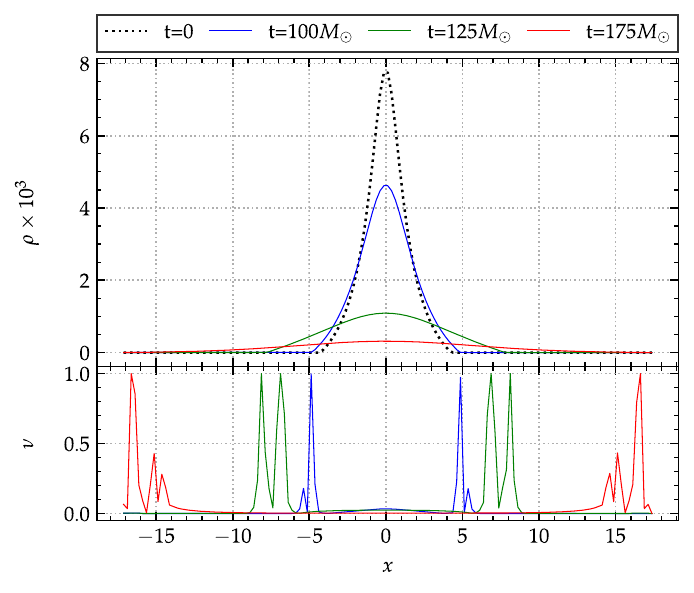}
   \caption{Migration test. Profiles of the rest-mass density (top panel) and of the corresponding entropy production (bottom panel) at different stages of the initial rapid expansion of the migrating TOV star, see \autoref{fig:rho_max_dyn_ideal_migration}. The solution is computed with the WENOZ reconstruction scheme and $n=128$.}
   \label{fig:1d_rho_nu_dyn_ideal_migration}
\end{figure}

\begin{figure}[h]
 \centering 
  \includegraphics[width=0.45\textwidth]{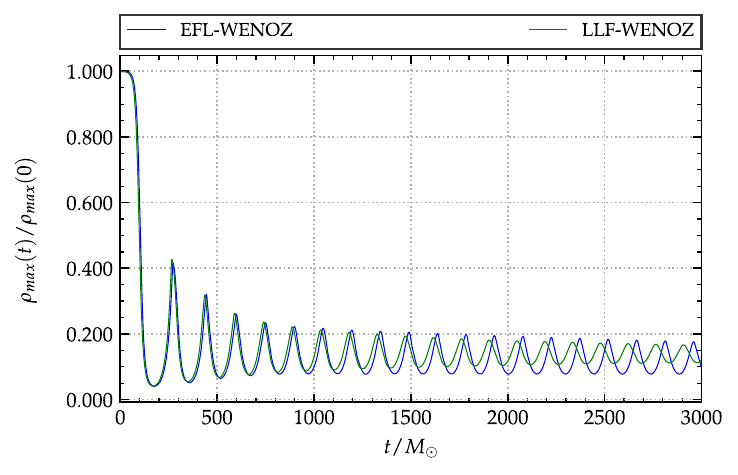}
   \caption{Evolution of the central rest-mass density for the migration test with $n=128$. The results of the EFL method are compared to the second-order scheme LLF-WENOZ used in \cite{Thierfelder:2011yi,Dietrich:2015iva}.}
   \label{fig:rho_max_dyn_ideal_migration}
\end{figure}

We start by inspecting the one-dimensional profiles of \autoref{fig:1d_rho_nu_dyn_ideal_migration} which depict the rest-mass density (top panel) and the entropy production function (bottom panel). The top panel shows the behaviour of the rest-mass density during the initial rapid expansion of the NS. The NS evolves from its initial unstable configuration at $t=0$ to $t=175 M_\odot$ where the central rest-mass density acquires its minimum value $\rho_c=3.139\times10^{-4}$, see \autoref{fig:rho_max_dyn_ideal_migration}. In just 0.8 ms the NS has expanded in such a degree that its central rest-mass density is about 4\% of its initial value. As can be seen on the bottom panel, the EFL method reacts extremely well to this rapid change of the location of the NS surface. The peaks on the profiles of the entropy production $\nu$ are localised correctly around the moving surface of the NS ensuring that the stable numerical flux is used there. This, in turn, guarantees that the numerical solution is free of spurious oscillations and long term stable.

We continue by checking the behaviour of the central rest-mass density with time. \autoref{fig:rho_max_dyn_ideal_migration} presents the evolution of the central rest-mass density for both the EFL and LLF-WENOZ method. The resulting oscillating behaviour is due to the strong non-linear pulsation of the originally unstable NS around a stable configuration. At every pulsation shock waves are formed that dissipate kinetic into thermal energy and gradually damp the oscillations. The results of the EFL method are comparable to the corresponding ones of the LLF-WENOZ scheme and the ones in \cite{Font:2001ew,CorderoCarrion:2008nf,Thierfelder:2011yi}.

\section{Conclusions}
\label{sec:conclusions}

The main purpose of the present work is to extend the use of the EFL method to scalar PDE systems. In addition, the EFL scheme is used to successfully simulate a couple of very demanding single NS spacetimes: the migrating and boosted TOV NS.

We extend the EFL method presented in \cite{Doulis:2022vkx} and show how to generalised it to scalar PDE systems in \autoref{sec:EFL_method} through the introduction of an auxiliary entropy functional. The resulting EFL scheme successfully passes in \autoref{sec:scalar_tests} a quite demanding set of scalar benchmark problems, with results comparable to the ones in the literature \cite{Toro:1999}.

In \autoref{sec:SRHD_tests} we revisit a couple of benchmark problems in special relativistic hydrodynamics that were also studied in \cite{Doulis:2022vkx}. Here, we focus on the study of some quantitative and qualitative properties of the entropy production function that provide a better understanding of the workings of the EFL method. Our results i) show the direct connection between the increasing use of the LO flux and the decreasing order of convergence and ii) demonstrate the extreme sensitive of the EFL method in recognising the slightest non-smoothness features present in our solutions.

Finally, the EFL method is tested in \autoref{sec:SNS} against three-dimensional general-relativistic non static/stationary single NS configurations. As expected, the EFL scheme accurately locates the surface of the NS, see \autoref{fig:1d_rho_nu_dyn_ideal_boost} and \autoref{fig:1d_rho_nu_dyn_ideal_migration}, and enables the use of the LO flux in this region while the interior remains mainly resolved by the HO flux. EFL simulations give results comparable to those obtained with standard WENO schemes \cite{Thierfelder:2011yi}. 

Naturally the question arises of whether the entropy pair introduced in \autoref{sec:EFL_method} and used to define the entropy of systems without EoS can be also used for systems with a well defined thermodynamic entropy like the ones of \autoref{sec:SRHD_tests}, \autoref{sec:SNS}. Such a possibility would highly simplify the use of the EFL scheme as the same entropy pair could be used for a wide variate of different conservation laws. The investigation of this possibility is currently in progress and results will be published as soon as available.

\begin{acknowledgements}

We would like to thank members of the Jena group for fruitful discussions and invaluable input. 
GD acknowledges  funding from the European High Performance Computing Joint Undertaking (JU) and Belgium, Czech Republic, France, Germany, Greece, Italy, Norway, and Spain under grant agreement No 101093441 (SPACE) and from the EU Horizon under ERC Starting Grant, grant agreement no.~BinGraSp-714626.  
SB acknowledges funding from the EU Horizon under
ERC Starting Grant, grant agreement no.~BinGraSp-714626, and 
ERC Consolidator Grant, grant agreement no.~InspiReM-101043372.
WT acknowledges support by the National Science Foundation under grant PHY-2136036.

Computations where performed on the ARA and DRACO clusters at
Friedrich Schiller University Jena. The ARA cluster is partially
funded by the DFG grants INST 275/334-1 FUGG and INST 275/363-1 FUGG,
and ERC Starting Grant, grant agreement no.~BinGraSp-714626.
The authors also gratefully acknowledge the Gauss Centre for Supercomputing
e.V. (\url{www.gauss-centre.eu}) for funding this project by providing
computing time on the GCS Supercomputer SuperMUC-NG at Leibniz
Supercomputing Centre (\url{www.lrz.de}) with the allocations
{\tt pn36ge} and {\tt pn36jo}.  

\end{acknowledgements}


%

\end{document}